\documentclass[reprint,amssymb,aps,prd]{revtex4-1}
\usepackage[utf8]{inputenc}
\usepackage{graphicx}
\usepackage{mathtools}
\usepackage[usenames,dvipsnames]{xcolor}
\usepackage{booktabs}
\usepackage[columnwise]{lineno}

\usepackage{enumitem}
\usepackage{ulem}
\usepackage{soul}
\usepackage{siunitx}
\DeclareSIUnit[number-unit-product = {}]\diopter{D}
\DeclareSIUnit[number-unit-product = {}]\parsec{Pc}

\usepackage{hyperref}
\hypersetup{
        colorlinks=true,
        citecolor=Turquoise,
        filecolor=blue,
        linkcolor=DarkOrchid,
        urlcolor=blue
        }

\begin{document}

\title{An analysis and visualization of the output mode-matching requirements for squeezing in Advanced LIGO and future gravitational wave detectors}
\author{Antonio Perreca}
\affiliation{University of Trento, Department of Physics, I-38123 Povo, Trento, Italy\\
INFN, Trento Institute for Fundamental Physics and Applications, I-38123 Povo, Trento, Italy}
\author{Aidan F. Brooks and Jonathan W. Richardson}
\affiliation{LIGO Laboratory, California Institute of Technology, Pasadena, CA 91125, USA}
\author{Daniel T{\"o}yr{\"a}}
\affiliation{Department of Quantum Science and Centre for Gravitational Physics, Australian National University, Canberra ACT 2600, Australia}
\author{Rory Smith}
\affiliation{OzGrav: The ARC Centre of Excellence for Gravitational-Wave Discovery, Monash University, VIC 3800, Australia}

\begin{abstract}
The sensitivity of ground-based gravitational-wave (GW) detectors will be improved in the future via the injection of frequency-dependent squeezed vacuum. The achievable improvement is ultimately limited by losses of the interferometer electromagnetic field that carries the GW signal. The analysis and reduction of optical loss in the GW signal chain will be critical for optimal squeezed light-enhanced interferometry. In this work we analyze a strategy for reducing output-side losses due to spatial mode mismatch between optical cavities with the use of adaptive optics. Our goal is not to design a detector from the top down, but rather to minimize losses within the current design. Accordingly, we consider actuation on optics already present and one transmissive optic to be added between the signal recycling mirror and the output mode cleaner. The results of our calculation show that adaptive mode-matching with the current Advanced LIGO design is a suitable strategy for loss reduction that provides less than 2\% mean output mode-matching loss. The range of actuation required is $+47\;\mu$D on SR3, $+140$~mD on OM1 and OM2, $+50$~mD on the SRM substrate, and $-50$~mD on the added new transmissive optic. These requirements are within the demonstrated ranges of real actuators in similar or identical configurations to the proposed implementation. We also present a novel technique that graphically illustrates the matching of interferometer modes and allows for a quantitative comparison of different combinations of actuators.
\end{abstract}

\maketitle

\section{Introduction}
\label{sec:intro}

The LIGO-Virgo Collaboration achieved the goal of detection of gravitational waves with the observation of GW150914~\cite{GW150914etal}. This was followed by the detection of several other binary black hole mergers \cite{bbh_mergers_aligo}. Additionally, gravitational waves from a binary neutron star (BNS) merger, GW170817, were observed with multiple coincident electromagnetic observations in August 2017 \cite{MMA_BNS:2017etal}. These observations mark the dawn of gravitational-wave astronomy, opening interstellar laboratories for tests of theories of matter and gravity in the strong regime.

\begin{figure}[t]
  \centering
  \includegraphics[width=\columnwidth]{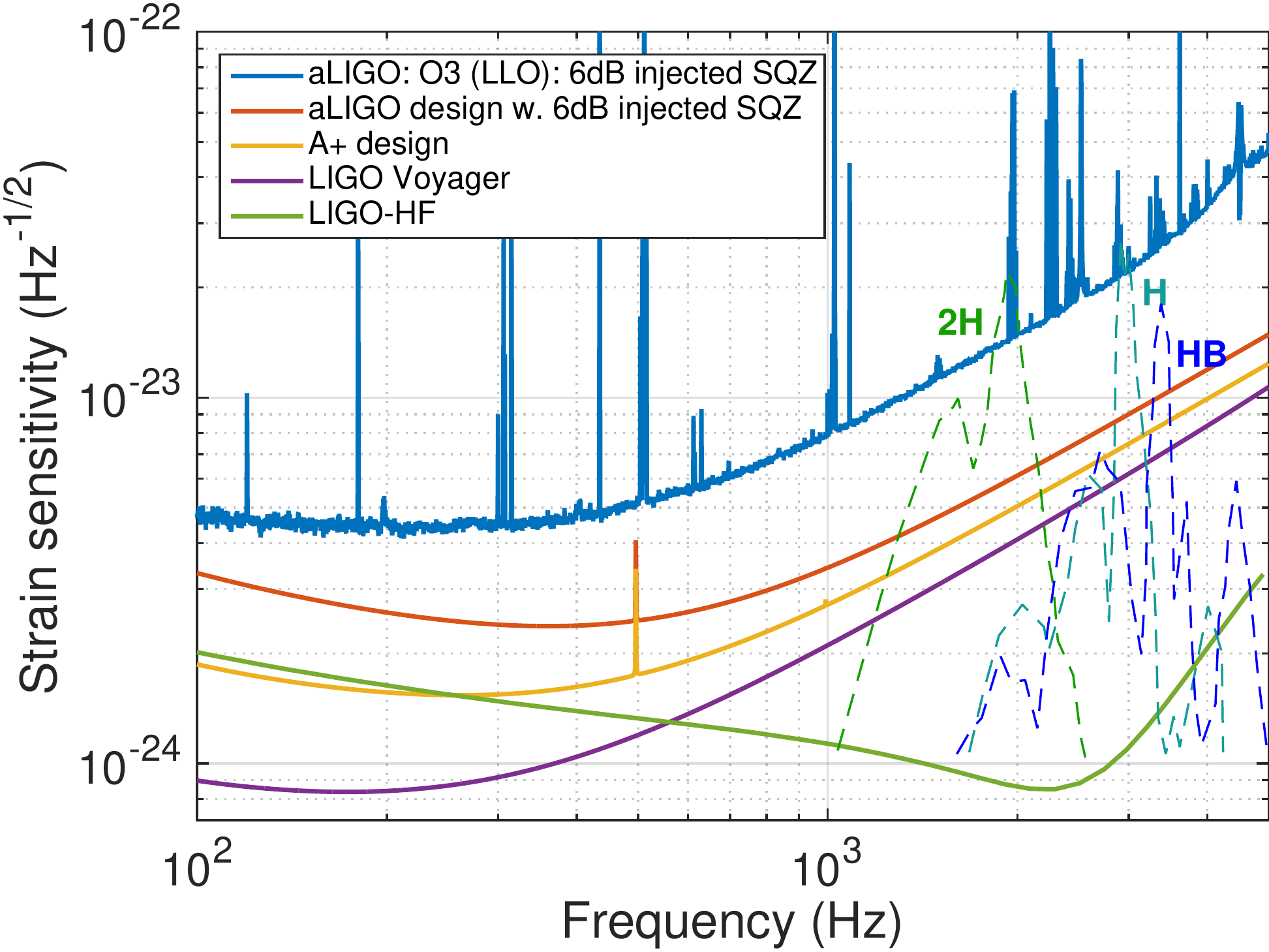}
  \caption{High-frequency strain sensitivity of current and proposed interferometer configurations. Blue curve: O3 Advanced LIGO (aLIGO) sensitivity with 6~dB of injected squeezing (3~dB observed, see \cite{Tse_2019}). Red curve: aLIGO design plus 6~dB of injected squeezing. Yellow curve: A+, a future upgrade to aLIGO with coating thermal noise reduced by a factor of 2 and 12~dB of injected frequency-dependent squeezing (reproduced from \cite{InstrumentScience:2018}). Purple curve: Voyager, a proposed upgrade of A+ with 15~dB of injected frequency-dependent squeezing and lower coating thermal noise (reproduced from Adhikari et al.~\cite{Voyager:2020}). Green curve: LIGO-HF, another proposed upgrade of A+ with optical parameters re-optimized for high-frequency sensitivity (reproduced from Martynov et al.~\cite{Martynov:2019_pub}). Also shown are merger waveforms for different NS equations of state (dashed lines). The simulations assume a reference BNS coalescence at 100 Mpc (courtesy J. Veitch and S. Vitale, adapted from \cite{PhysRevD.88.044042}).}
  \label{fig:strain_sens}
\end{figure}

The strain sensitivity of the LIGO detectors~\cite{aLIGO_Overview}, shown in Figure \ref{fig:strain_sens}, is limited above approximately \SI{200}{\hertz} by quantum noise (vacuum fluctuations) in the form of shot noise. For full details on LIGO noise, see Martynov et al.~\cite{Martynov_noise:15etal}. The high-frequency sensitivity is of interest because one of the many goals of gravitational-wave astronomy is to observe the merger phase of a binary neutron star (BNS) merger, thereby gaining insight into the neutron-star (NS) equation of state \cite{NSEOS_2009,PhysRevD.88.044042}. The dynamics of this merger phase are typically encoded in the quantum-noise-limited frequency range between \SI{1.5}{} and \SI{5}{\kilo\hertz}. For example, the gravitational wave signal from GW170817~\cite{PhysRevLett.119.161101etal}, a characteristic chirp increasing in frequency, fell below the LIGO noise floor around \SI{400}{\hertz} and thus provided limited information about the merger phase and NS equation of state.

At this time, the LIGO detectors are operating with a neutron-star-neutron-star (NSNS) sensitivity of around 115~MPc at LIGO-Hanford (LHO) and 140~MPc at LIGO-Livingston (LLO). This is not yet at the design sensitivity, approximately 190~MPc, and is largely limited by technical noises at low frequencies (below 100~Hz) and shot noise at higher frequencies \cite{Martynov_noise:15etal}. We expect to reduce these noise sources and achieve the design sensitivity within a few years \cite{ObservingScenarios2018etal}. The high-frequency sensitivity will be improved using the technology known as squeezing \cite{Cav1980,CaSc1985}. A significant improvement in the high-frequency sensitivity brings with it a commensurate improvement in our ability to measure the NS equation of state during a BNS merger.

\begin{figure}[t]
  \centering
  \includegraphics[width=\columnwidth]{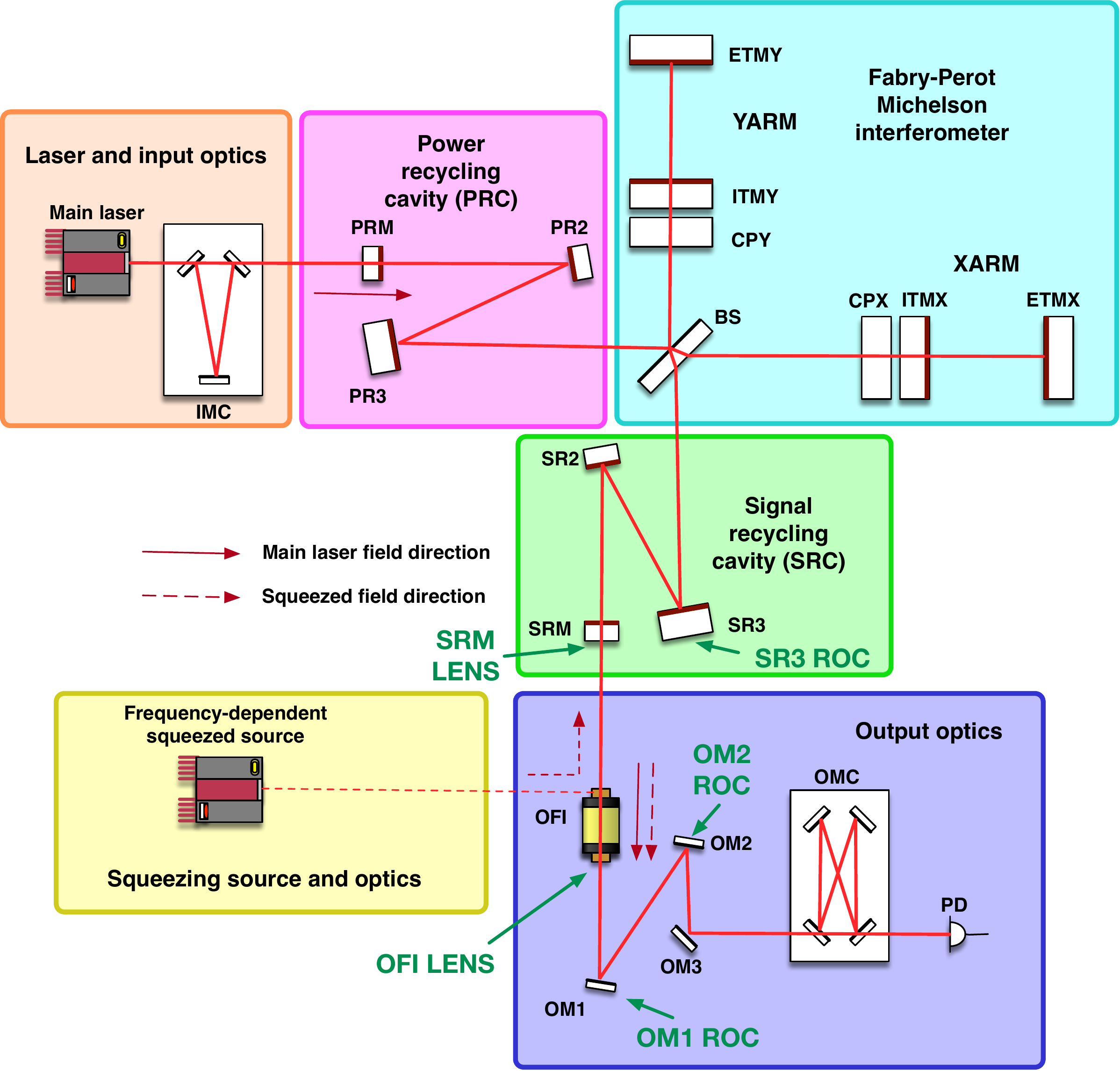}
  \caption{Advanced LIGO interferometer with a frequency dependent squeezed source. The squeezed vacuum field (dashed line) is injected via the output Faraday isolator (OFI), a bi-directional coupler. Although active wavefront control is also required within the frequency dependent squeezed source, it is not discussed in this manuscript (see \cite{FQ_dep_SQZ} for more details). The main modules of the LIGO detector are highlighted by colored boxes. Some optics have been omitted for clarity. The locations of current and possible future output mode-matching actuators (SR3 ROC, SRM LENS, OFI LENS, OM1 ROC and OM2 ROC) are highlighted.}
   \label{fig:IFO}
\end{figure}

Squeezing involves the preparation of a vacuum state in which fluctuations (quantum noise), initially distributed uniformly between amplitude and phase quadratures, are redistributed so that they are suppressed in one quadrature and amplified in the other. Traditionally, a squeezed vacuum state is prepared with an optical parametric oscillator such that vacuum fluctuations are redistributed from the readout quadrature (of phase-amplitude space) to the orthogonal quadrature~\cite{Kirk:2004}. As illustrated in Figure~\ref{fig:IFO}, one injects this squeezed field into the interferometer via a directional port (Faraday isolator) close to the output of the interferometer~\cite{LIGOsq13}. The squeezed light propagates through the interferometer, eventually reaching the output photodetectors and reducing the shot noise below the standard vacuum level. In the last few years, the technology for generating squeezed light has reached maturity and its performance is constantly improving \cite{Kirk:2004, Henning:PRL2006, Go:40m, Vahlbruch07, GEO:Squeezing, LIGOsq13}. Advanced LIGO is now routinely operating with 2.7~dB of observed squeezing, and has demonstrated performance as high as 3.2~dB \cite{Tse_2019}.

The injection of phase-squeezed (or frequency-independent-squeezed) light adds additional amplitude noise which beats with the interferometer electric field. This applies a force noise to the optics via radiation pressure, increasing the displacement noise at low frequencies. If the squeezed field is first reflected off a detuned filter cavity whose cavity pole is near the cross-over frequency of radiation pressure and shot noise ($\sim$\SI{100}{\hertz} for Advanced LIGO), the squeezed quadrature will be rotated in a frequency-dependent way. This allows one to achieve amplitude-quadrature squeezing at low frequencies and phase-quadrature squeezing at high frequencies, thereby reducing the quantum noise of the interferometer at all frequencies~\cite{Evans_FC:13, Khalili:10}. Frequency-dependent squeezing has been experimentally demonstrated~\cite{FQ_dep_SQZ} and is planned for future installation in LIGO.

\begin{figure}[t]
  \centering
  \includegraphics[width=\columnwidth]{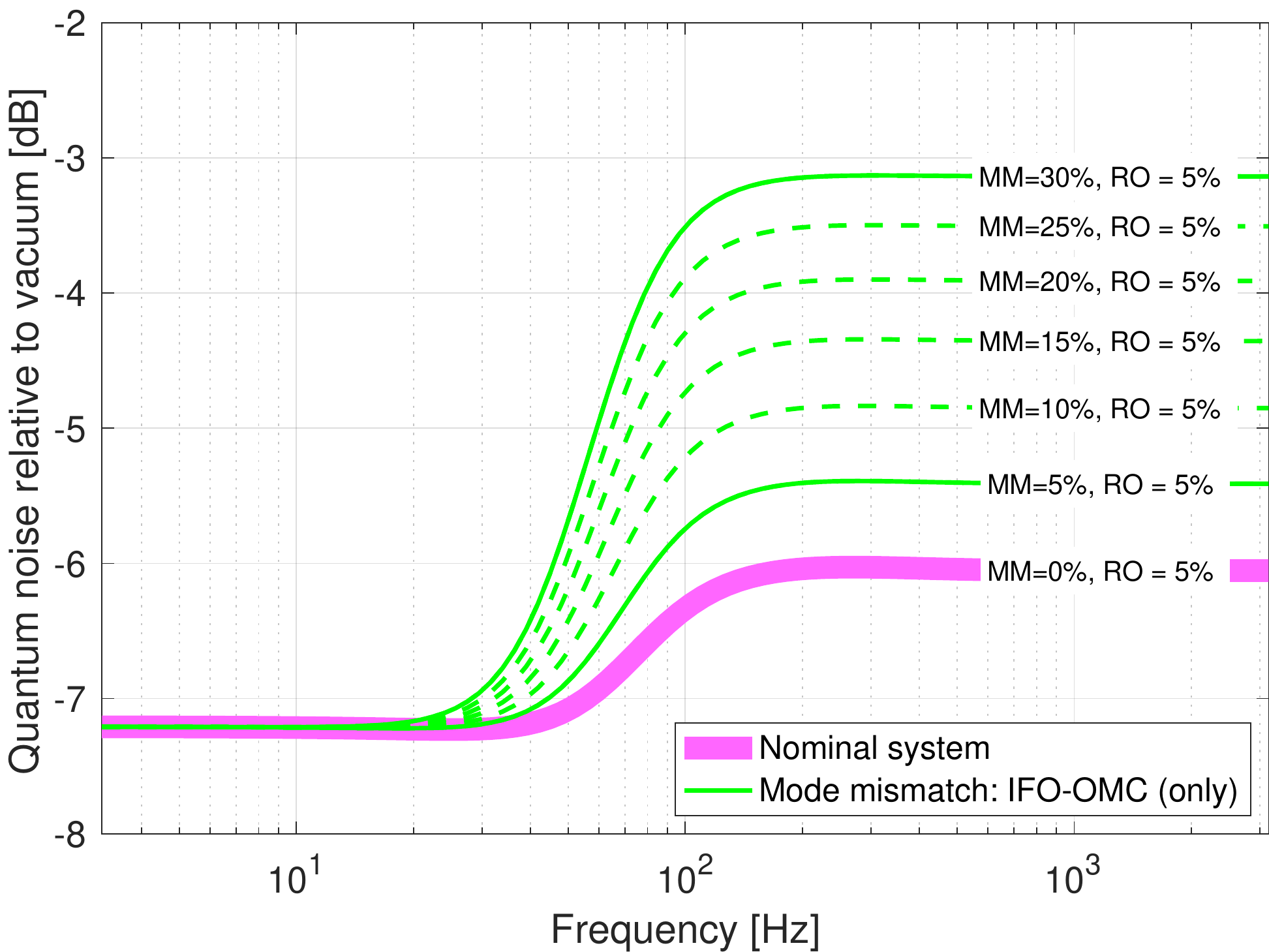}
  \caption{Impact of output mode-matching loss on frequency-dependent squeezing. The interferometer quantum noise, relative to unsqueezed vacuum, is shown for different mode-matching losses between the interferometer and OMC (green curves). Poor mode-matching causes a dramatic degradation of the quantum noise reduction at frequencies above the filter cavilty pole. The model parameters are the same as in Table 2 of \cite{FQ_dep_SQZ}, but with the filter cavity length extended from 16 to 100~m and the phase noise reduced to 5~mrad to reflect recent advances~\cite{2016_Oelker_PhaseNoise}. The mode mismatch (MM) is indicated for each curve. All cases (green and magenta) assume an additional 5\% readout (RO) loss not associated with mode-matching (e.g., quantum efficiency of photodiodes).}
   \label{fig:SQZ_v_FREQ}
\end{figure}

Although the injected level of squeezing can be high, the observed level of squeezing in a real interferometer will be limited by losses in the interferometer and the quadrature fluctuations of the input squeezed field. Losses partially mix the squeezed vacuum state with unsqueezed vacuum. Losses arise from scattering, reflections from optics, photodetector quantum efficiency, and mode mismatches among cavities. In Enhanced LIGO, the dominant optical losses (25\% $\pm$ 5\%) were caused by mode-mismatches between cavities due to variation of the optics parameters from their nominal values \cite{LIGOsq13}. For squeezing, mode-matching losses occur when coupling the squeezed field into the interferometer and also when coupling the interferometer field through the signal-recycling cavity (SRC) and output mode cleaner (OMC), as illustrated in Figure~\ref{fig:IFO} and described in \cite{SQZ_2014_Oelker}. The current output mode-matching loss is estimated to be at least 10\% for LIGO Livingston \cite{aLOG_MM_LLO}, making it a limiting source of squeezing loss. Figure~\ref{fig:SQZ_v_FREQ} illustrates the impact output mode-matching loss will have on frequency-dependent squeezing. Poor mode-matching will result in a dramatic degradation of the quantum noise reduction at frequencies above the filter cavity pole.

In this paper, we present a study of the output mode-matching between the interferometer and the OMC. Oelker et. al. \cite{SQZ_2014_Oelker} have found that -8 to -10~dB of squeezing is possible when quadrature fluctuations are reduced to a few milliradians and the {\it total} losses are limited to 10\% to 15\%, a level projected as achievable in Advanced LIGO in the near future. In order to achieve this total loss, the output mode-matching loss from the interferometer to the OMC must be reduced to 2 to 3\%. Therefore, this paper aims to determine the active wavefront control requirements necessary to achieve better than 98\% output mode-matching.

By its very nature, mode-mismatch in an interferometer occurs due to deviations from the nominal design (which assumes perfect mode-matching). For example, tolerances on the polishing of optics admit a range of possible radii of curvature, and optics can only be placed inside the interferometer to certain precision. In general, deviations from design can cause mode-mismatch at any spatial order. However, the known sources described above induce a radius of curvature (ROC) mismatch between cavities, correctable with spherical lensing actuation. Moreover, we will show that these sources alone can fully account for the current output mode-mismatch (see \S\ref{sec:MMopt}). Thus it is reasonable to assume that the mode-mismatch losses are dominated by low-order effects and to design an actuation strategy targeting them. However, it is not known how much residual mismatch will remain after the low-order terms have been mitigated.

The paper is organized as follows. In \S\ref{sec:actuators} we discuss the interferometer modes that must be matched as well as the actuation locations that are accessible. In \S\ref{sec:ws} we describe a phase space that graphically illustrates mode-matching, aids building an intuitive picture of mode-overlap, and allows for a quantitative comparison of different combinations of actuators. In \S\ref{sec:act_strategy} we propose an actuation strategy developed using these visualizations, and then confirm its mode-matching capability using a full multi-mode statistical model of the interferometer including all optical tolerances. Concluding remarks are presented in \S\ref{sec:conclusion}.


\section{Interferometer modes and actuators}
\label{sec:actuators}

To examine the interferometer mode-matching, we first identify the modes in question and the locations of existing and potential radius-of-curvature (ROC) actuators.

\subsection{Interferometer modes}
\label{sec:ifo_modes}

Within a dual-recycled Fabry-Perot Michelson interferometer with frequency-dependent squeezing there are eight fundamental (Gaussian) optical modes that, ideally, are perfectly matched to each other and the input laser beam: input mode cleaner (IMC), power recycling cavity (PRC), two Fabry-Perot arm cavities (XARM, YARM), signal recycling cavity (SRC), output mode cleaner (OMC), squeezer, and filter cavity (FC). For the purposes of this discussion, we ignore the input modes (IMC and PRC), and assume that the 4~km Fabry-Perot XARM and YARM modes are identical (that is, we assume differential mode-mismatch is corrected by the existing thermal compensation system). We represent the common-arm Fabry-Perot mode as the ARM mode. Additionally, we ignore the matching of the squeezer and FC modes to the interferometer as this is considered elsewhere \cite{FQ_dep_SQZ}. This leaves us with the following relevant interferometer modes:

\begin{enumerate}
\item Signal recycling cavity (SRC)
\item Common-arm Fabry-Perot (ARM)
\item Output mode cleaner (OMC)
\end{enumerate}

\subsection{Adaptive optic actuators}
\label{sec:actuator_designs}

Figure~\ref{fig:IFO} shows the layout of Advanced LIGO, which includes an existing set of mode-matching actuators known as the thermal compensation system (TCS) \cite{Brooks:16}. Briefly, ring heaters encircle each of the four test masses (ITMX, ETMX, ITMY, and ETMY) to adjust their ROC and $\rm CO_2$ laser actuators heat the compensation plates (CPX and CPY) located outside of the Fabry-Perot arms to induce thermal lenses. For full details, see Brooks et al.~\cite{Brooks:16}. These existing TCS actuators are degenerate with respect to the PRC and SRC. That is, one cannot actuate with the TCS actuators to affect the SRC mode without also affecting the PRC mode.

Currently, the TCS actuators serve to correct dynamic changes in the ITM and ETM surface curvatures and substrate lenses and are also used to remove static lenses in the ITM substrates (particularly differential lenses). The remaining TCS degree of freedom is the common recycling cavity lens (CO2COM),

\begin{equation}
\mathrm{CO2COM} = \frac{\mathrm{CO2X} + \mathrm{CO2Y}}{2},
\end{equation}

\noindent where CO2X and CO2Y are the lenses induced in CPX and CPY, respectively. The CO2COM lens is used to optimize the PRC-ARM coupling.

On the output side of the interferometer, the other actuators shown in Figure~\ref{fig:IFO} include: an SR3 ROC actuator that allows limited control over the ROC of the SR3 optic, a tunable lens in the SRM substrate and/or a new transmissive optic just after the SRM (OFI LENS), and OM1 and OM2 ROC actuators. OM3 is not suitable for use as an actuator because the angle of incidence of the laser beam on that optic is large enough to create significant astigmatism when spherical changes are made to the OM3 ROC. Of these actuators, only the SR3 ROC actuator currently exists in Advanced LIGO. The details of these actuator designs are discussed in Appendix~\ref{sec:actuators_physical}.


\section{Mode-matching visualization: WS phase space}
\label{sec:ws}

In this section, we describe a novel graphical technique for visualizing mode-matching between different interferometer modes, expanding on earlier work \cite{Brooks_LVC:15}. We designate this the ``WS phase space'', or simply ``WS space.'' This provides a visual representation of the magnitude and relative actuation phase of the actuators on the previously discussed modes.

\subsection{WS phase space overview}

We construct a two-dimensional phase space that is spanned by {\it beam size}, $W$, (x-axis) and {\it defocus}, $S$, (y-axis). Specifically, $W$ is defined as the $1/e^2$ radius of the beam intensity profile and $S$ as the inverse of the radius of curvature. The sign of $S$ is defined such that a beam that is {\it converging} to a waist is defined to have a negative defocus, while a beam that is {\it diverging} away from a waist has a positive defocus.

\begin{figure*}[t]
  \begin{minipage}[l]{\columnwidth}
    \centering
    \includegraphics[trim=40 0 70 0, clip, width=\columnwidth]{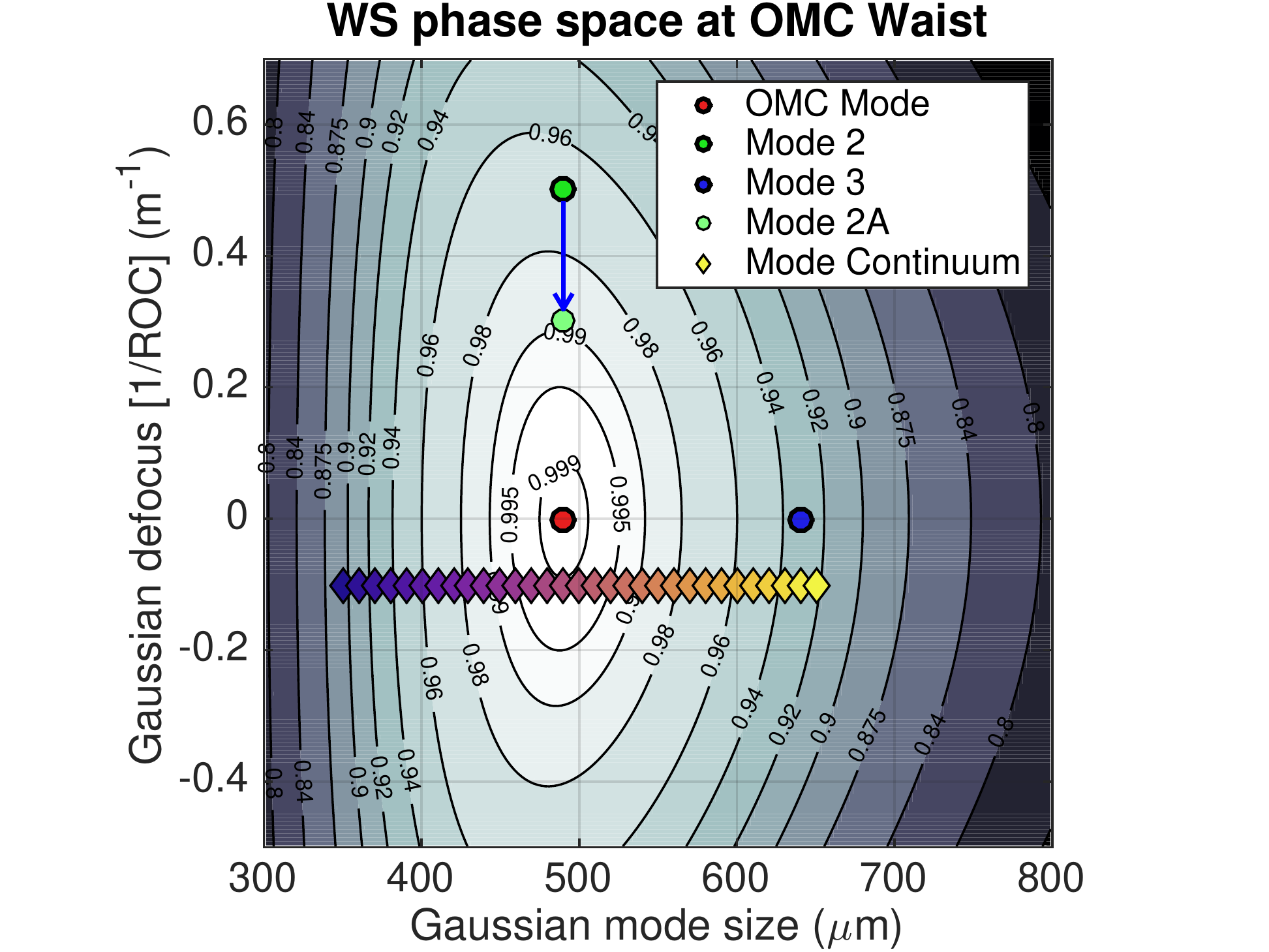}
  \end{minipage}
  \hfill{}
  \begin{minipage}[r]{\columnwidth}
    \centering
    \includegraphics[trim=40 0 70 0, clip, width=\columnwidth]{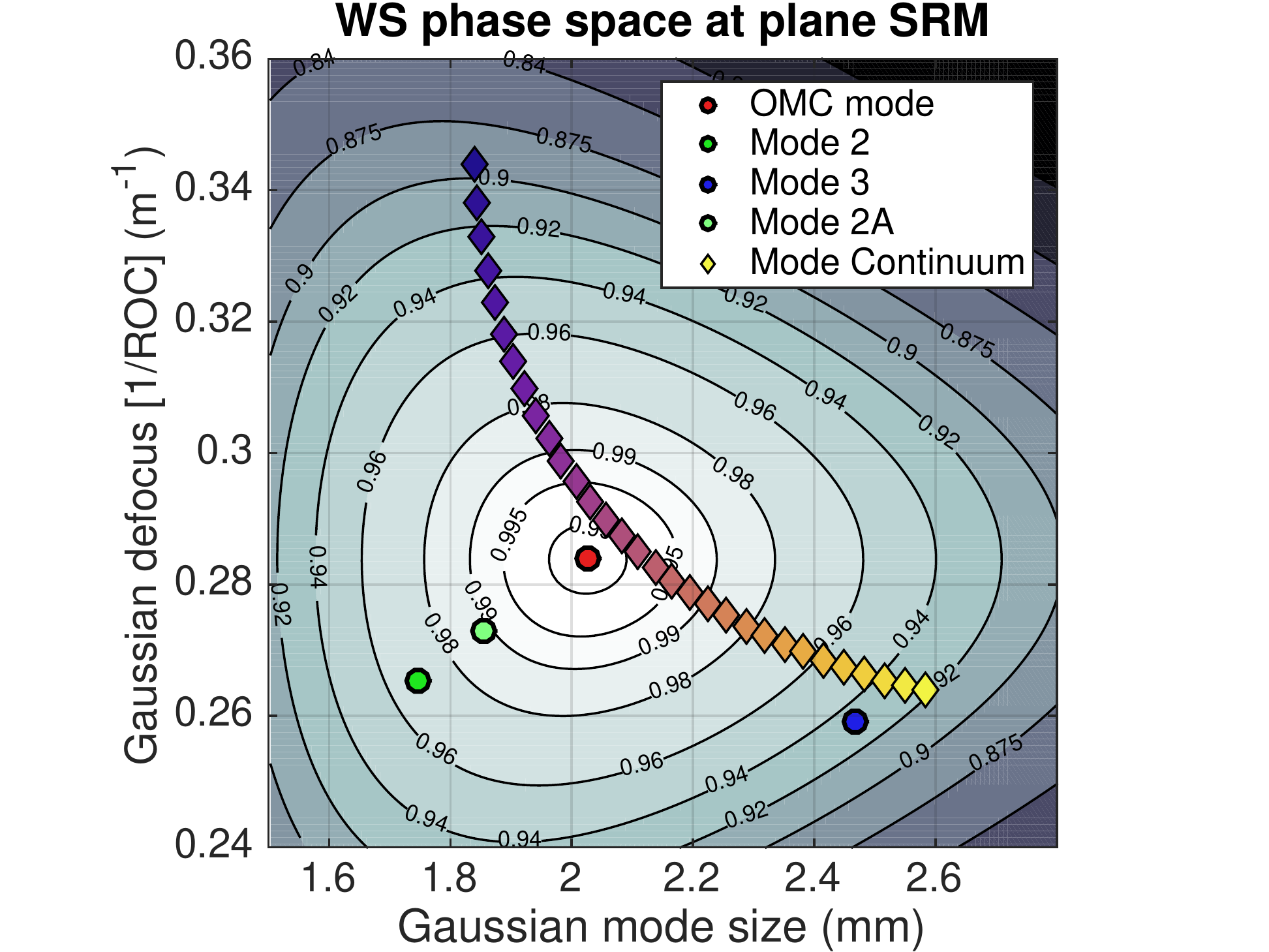}
  \end{minipage}
  \caption{Left: The WS phase space where the primary mode $\left(W_{P}, S_{P}\right)$ is the OMC mode at the plane of the OMC waist. Mode~2 has the same beam size as the primary mode but different defocus. Mode~3 has the same defocus as the primary mode but different beam size. Mode~2A represents the actuation of Mode~2 by a positive lens of defocus 0.2~D applied at the plane of the OMC waist. Finally, a continuum of modes is shown for comparison to different longitudinal planes (see right panel). Right: The WS phase space at the plane of the SRM. All modes from the left panel have been propagated to this plane to illustrate their evolution through WS phase space (see \S\ref{sec:propagation}). The primary mode is still the OMC mode.}
  \label{fig:MM_space_example}
\end{figure*}

A purely Gaussian mode at a longitudinal plane, $z$, is fully defined by its beam size, $W$, and defocus, $S$. Such a mode can be represented within this phase space as a single point with those values as coordinates. In terms of the complex beam parameter, $q$, the Gaussian mode is
\begin{equation}
\frac{1}{q} = S - \frac{i\, \lambda}{\pi \, W^2}.
\end{equation}
All additional Gaussian modes, when propagated to the same longitudinal plane, can also be represented within this phase space and compared to the primary mode. Ignoring higher-order spatial modes, if two modes have the same beam size and defocus, then they have 100\% mode overlap and occupy the same location in this space. If they differ in size and/or defocus, then they have less than 100\% overlap and occupy different locations in WS space. This space is illustrated in the left panel of Figure~\ref{fig:MM_space_example}, which shows the overlap with the aLIGO OMC mode at the location of the OMC waist.

For the primary mode $\left(W_{P}, S_{P}\right)$ under consideration in the WS space (the red point at the center of the phase space in Figure~\ref{fig:MM_space_example}, left panel), we determine the mode-overlap with every other point $\left(W,S\right)$ in the space as
\begin{align}
\mathrm{OL}\left(W,S\right) & = & \iint E\left(W, S\right) \, E\left(W_{P}, S_{P}\right)^* dx\, dy \, \times \nonumber \\
     & & \iint E\left(W, S\right)^* \, E\left(W_{P}, S_{P}\right) dx\, dy\,,
\label{eqn:OL}
\end{align}
where $E\left(W, S\right)$ is given by
\begin{eqnarray}
E\left(W, S\right) = \sqrt{\frac{2}{\pi}} \, \frac{1}{W} \, \exp \left(- i k \frac{x^2 + y^2}{2 \, q(S,W)} \right) \nonumber \\
&&
\end{eqnarray}
which has unit normalization. This allows us to construct a set of iso-overlap contours centered around $\left(W_{P}, S_{P}\right)$, as illustrated in the left panel of Figure~\ref{fig:MM_space_example}. In this figure, we have plotted several additional modes for illustration. For example, ``Mode~2'' and ``Mode~3'' have greater defocus and beam size, respectively. The resulting overlap of Mode~2 or Mode~3 with the primary mode is easily inferred from the nearest iso-overlap contours.

Once this is done, and other modes are plotted within this space, the overlap of all modes with the primary mode is apparent. Additionally, we can visually interpret the gradient of the mode-overlap as a function of $W$ and $S$ by the density (and values) of the contours. The mode-mismatch loss between any point and the primary mode is readily inferred as $L = 1 - OL$. In constructing and interpreting these plots, the following rules apply:
\begin{enumerate}
\item All modes must be represented at the same longitudinal plane in the optical chain.
\item The contours around the primary mode only convey information about the overlap of the primary mode with all other points. For example, with contours centered around the primary mode, mode $A$, and with modes $B$ and $C$ represented within that space, this representation does {\it not} convey the overlap between $B$ and $C$, even if they happen to have the same overlap value with the primary mode.
\end{enumerate}

\subsection{Actuation on modes}
\label{sec:mode_actuation}

We now consider visualization of actuation on a mode. In a real interferometer, we have no simple means to directly change the beam size of a mode while preserving its total power (that is, we cannot use apertures or apodized masks to change the beam size without reducing the overall power in the mode). Consider Mode~3 in Figure~\ref{fig:MM_space_example}, left panel, which is matched in defocus but not in beam size. We cannot improve the overlap of Mode~3 with the primary mode by actuation at this longitudinal plane, ($z_{\mathrm{OMC }}$). However, we have a straightforward means of changing the defocus of a mode: namely, adding lensing to that mode using, for example, an actuator similar to one of those described in \S\ref{sec:actuator_designs}. Consider Mode~2 in Figure~\ref{fig:MM_space_example}, left panel, which is instead matched in beam size but not in defocus. The mode-matching to the primary mode is observed to improve when we apply +0.2 diopters of lens power (Mode 2A in Figure~\ref{fig:MM_space_example}, left panel).

To expand upon this idea, consider the interferometer modes defined in \S\ref{sec:ifo_modes} propagated to the location of one of our actuators, for example, the longitudinal plane immediately following the SRM. Any defocus, $S_{\rm SRM}$, applied by that actuator will simply be added to the defocus of the interferometer modes. In an optical ABCD matrix formalism, this is equivalent to adding the following matrix at the plane $z_{\rm SRM}$:
\begin{equation}
\left| \begin{array}{cc} A & B \\ C & D \end{array} \right| = \left| \begin{array}{cc} 1 & 0 \\ -S_{\rm SRM} & 1 \end{array} \right| \;.
\end{equation}
This matrix applied to the complex beam parameter yields
\begin{eqnarray}
\frac{1}{q_2} & = & \frac{C\, q_1 + D}{A \, q_1 + B} \nonumber \\
 & = & C + \frac{1}{q_1} \nonumber \\
 & = & \left(S - S_{\rm SRM}\right) - \frac{i\, \lambda}{\pi \, W^2}\;,
\end{eqnarray}
implying the new defocus of the mode is $S - S_{\rm SRM}$.

Within the WS space represented at $z_{\rm SRM}$, all modes that {\it interact with that lens} will shift by $S_{\rm SRM}$ diopters. Just as the $C$ term in an ABCD matrix equals $-1/f$ for a standard lens, a positive thermal lens will reduce the defocus of a beam, while a negative lens will increase it. If the ARM mode is propagated from the ITMs to this plane, the last optical effect it experiences is this lens, and hence it accumulates this defocus change. The OMC mode, on the other hand, is propagated in the opposite direction (upstream) from the OMC to the SRM anti-reflective (AR) surface. It does not interact with actuator. Therefore, when this actuation is represented within the WS space, the ARM mode will move relative to the OMC mode, causing their mode-matching to change.

\subsection{Propagation to different longitudinal planes}
\label{sec:propagation}

We can represent the mode-overlap at {\it any} longitudinal plane of an unapertured optical system. The overlap between two Gaussian modes is independent of the longitudinal plane at which it is determined (this follows from the orthonormality of Hermite-Gauss modes, which is independent of longitudinal coordinate). Hence, as the longitudinal plane of the WS space is changed, the positions of two points in the space evolve such that their mode-overlap remains unchanged.

The propagation between two longitudinal planes separated by $\Delta z$ is governed by the ABCD matrix
\begin{equation}
\left| \begin{array}{cc} A & B \\ C & D \end{array} \right| = \left| \begin{array}{cc} 1 & \Delta z \\ 0 & 1 \end{array} \right| \;.
\end{equation}
This matrix, applied to the complex beam parameter at longitudinal plane~1, yields for longitudinal plane~2
\begin{eqnarray}
\frac{1}{q_2} & = & \frac{C\, q_1 + D}{A \, q_1 + B} \nonumber \\
& = & \frac{1}{q_1 + B} \nonumber \\
& = & \left(\frac{\kappa - 1 - S \, \Delta z}{\kappa \, \Delta z}\right) - \frac{i \lambda}{\pi \left(W \sqrt{\kappa}\right)^2} \;,
\end{eqnarray}
where
\begin{equation}
\kappa = 1 + 2 \, S \, \Delta z + \left[S^2 + \left(\frac{\lambda}{\pi W^2}\right)^2\right] \left(\Delta z\right)^2 \;.
\end{equation}
The result is a contour plot which nonlinearly distorts as it is propagated through an optical system, but which retains a one-to-one correspondence between the initial and final longitudinal planes. This is illustrated in the right panel of Figure~\ref{fig:MM_space_example}, which shows the WS space from the left panel (the OMC waist) propagated to a new longitudinal plane (denoted SRM).

In the context of Advanced LIGO, this can be helpful when visualizing the interferometer modes at different locations within the interferometer (e.g., at the beamsplitter, or at the nominal waist location of the OMC). Modes can be propagated backwards as well as forwards, as determined by the sign of $\Delta z$. Hence, we can propagate the OMC mode back to the beamsplitter just as easily as we can propagate the ARM mode to the OMC.

\subsection{Multiple actuators and Gouy phase}
\label{sec:gouy_phase}

One convenient feature of this representation is the ability to easily illustrate the effect of actuators at longitudinal planes other than where they are applied, as illustrated by the comparison of the left and right panels of Figure~\ref{fig:MM_space_example}. If we combine the effects of actuators at different longitudinal planes, we visualize areas or regions in the phase space that are accessible with these actuators. We now examine what determines the accessible region.

As shown in \S\ref{sec:mode_actuation}, the effect of an ROC actuator on a Gaussian mode specified by the parameters $\left(W, S\right)$ is to shift the defocus by $\Delta S$. Anderson \cite{Anderson:84} demonstrates that, in terms of the original Gaussian mode, $E(W,S)$, the actuation can be {\it approximately} represented as the addition of a purely imaginary Laguerre-Gauss 1-0 mode (LG10). That is, the actuated field can be expressed as
\begin{eqnarray}
E(W, S + \Delta S) & \simeq & \left(1-\frac{a^2}{2}\right)E(W,S) + i a \,E_{\rm LG10}(W,S) \nonumber \\
\label{eq:mode_decomp}
\end{eqnarray}
where the amplitude of the LG10 field is
\begin{equation}
a = \frac{\pi\,W^2}{\lambda} \, \frac{\Delta S}{2} \;.
\label{eq:LG10_mag}
\end{equation}
This linearized approximation is valid for small actuation coefficients, $a << 1$, or $\Delta S << 2 \lambda / \pi W^2$. As the following interpretations will rely on this decomposition, they apply only in the small-actuation regime. For larger actuations, the linearized LG10 approximation breaks down due to the coupling of additional higher-order modes.

\subsubsection{Actuator orthogonality}
\label{sec:orthogonality}

Suppose the Gaussian mode, propagated between longitudinal planes~1 and 2, accumulates a Gouy phase shift of $\phi_G$. The LG10 mode created by an actuator will accumulate a larger phase shift,
\begin{eqnarray}
\phi_{\rm HOM} &=& (2p + |l| + 1) \, \phi_G \nonumber \\
&=& 3 \, \phi_G \;,
\end{eqnarray}
where $p=1$ and $l=0$ are the radial and azimuthal orders, respectively. Thus, propagated along the longitudinal axis, the LG10 mode advances in phase by $2 \, \phi_G$ relative to the co-propagating Gaussian mode.

\begin{figure}[t]
  \centering
  \includegraphics[width=\columnwidth]{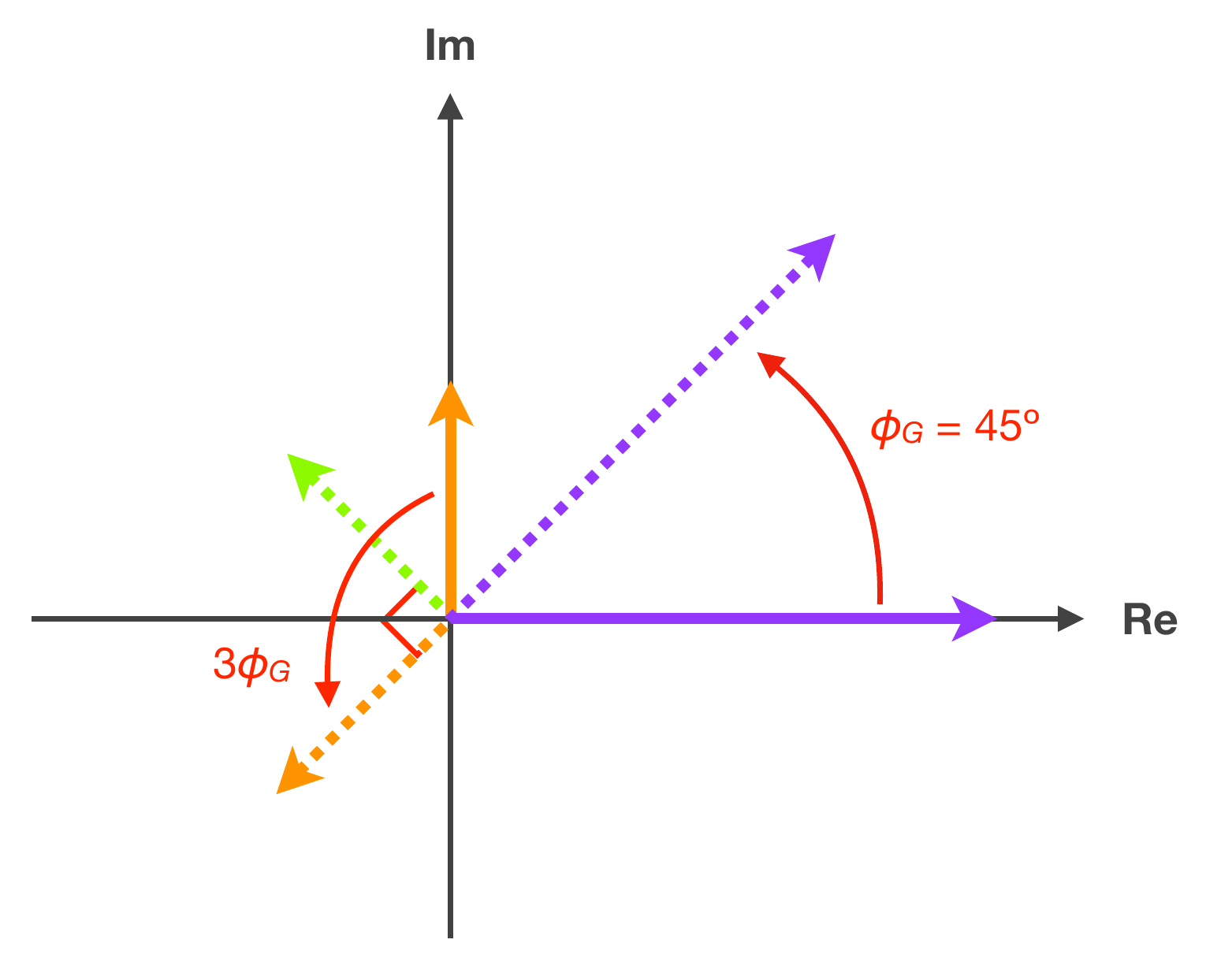}
  \caption{Rotation of complex mode amplitudes under longitudinal propagation. The solid lines show a Gaussian mode (purple) and an LG10 mode created by an ROC actuator (orange) at longitudinal plane~1. The LG10 mode is generated at $90^{\circ}$ from the Gaussian mode, as required by Equation~\ref{eq:mode_decomp}. The dashed lines show these modes propagated to longitudinal plane~2, in the case that the Gouy phase separation is $45^{\circ}$. A second actuator at longitudinal plane~2, represented by the green curve, generates an LG10 mode at $90^{\circ}$ from the propagated Gaussian mode. The relative phase of the two LG10 modes, compared at the same longitudinal plane, is thus $90^{\circ}$, indicating that the two actuators actuate on orthogonal quadratures.}
  \label{fig:complex_plane}
\end{figure}

Figure~\ref{fig:complex_plane} illustrates this effect for the special case that $\phi_G = 45^{\circ}$ between the two planes. The solid lines show the complex amplitudes of the Gaussian mode (purple) and the LG10 mode created by an actuator (orange) at longitudinal plane~1. The LG10 mode is generated at $90^{\circ}$ from the Gaussian mode, as required by Equation~\ref{eq:mode_decomp}. The dashed lines show these modes propagated to longitudinal plane~2. A second actuator at longitudinal plane~2, represented by the green curve, will generate an LG10 mode at $90^{\circ}$ from the propagated Gaussian mode. The relative phase of the two LG10 modes, compared at the same longitudinal plane, is $90^{\circ}$. In this case, the two actuators actuate on {\it orthogonal quadratures}.

From the above geometrical representation, it is clear that for {\it any} two ROC actuators, the angle between their LG10 modes in complex phase space (propagated to the same longitudinal plane) is $2 \, \phi_G$. In general, the degree of orthogonality of two ROC actuators is then
\begin{equation}
\gamma = \left| \sin\left( 2 \, \phi_G \right) \right| \;,
\label{eq:gamma}
\end{equation}
which depends only on the Gouy phase separation of the two longitudinal planes. A $\gamma$ value of 1 corresponds to orthogonality and a value of 0 to complete degeneracy. Table~\ref{Tab:gamma_matrix} lists the {\it cumulative} Gouy phase at each of the actuator locations discussed in \S\ref{sec:actuator_designs}, as well as the $\gamma$-values for different pairings of actuators.

\begin{table}
\centering
\resizebox{\columnwidth}{!}{
\begin{tabular}{ | l | c | c | c | c| c |}
    \hline
&  OM2&  OM1&  FI&  SRM&  SR3
  \\ \hline\hline
{\bf Beam size (mm)}  & 0.70 & 0.67 & 2.1 & 2.8 & 64.3\\ \hline
{\bf Gouy phase (deg)}  & 160.0 & 75.2 & 15.2 & 11.9 & 0.1\\ \hline
 \hline
    {\bf Orthogonality, $\gamma$ }  & & & & & \\ \hline
OM2  & 0  & 0.18 & {\bf0.94} & {\bf0.90} & 0.65 \\ \hline
OM1  & 0.18 & 0  & {\bf0.87} & {\bf0.80} & 0.50 \\ \hline
FI  & {\bf0.94} & {\bf0.87} & 0  & 0.11 & 0.50 \\ \hline
SRM  & {\bf0.90} & {\bf0.80} & 0.11 & 0  & 0.40 \\ \hline
SR3  & 0.65 & 0.50 & 0.50 & 0.40 & 0  \\ \hline
    \hline
 \end{tabular}}
\caption{The  beam  size  and  accumulated  Gouy  phase  of the interferometer TEM00 mode as it propagates from the ITM to the OMC. The gamma value for different actuator combinations is also shown, where $\gamma = 0$ indicates complete degeneracy and $\gamma=1$ indicates orthogonality.}
\label{Tab:gamma_matrix}
\end{table}

\subsubsection{Phase-space area}

The amplitude vectors of the LG10 modes created by two actuators (e.g., the dashed orange and green lines in Figure~\ref{fig:complex_plane}) trace out a {\it parallelogram area} in complex phase space,
\begin{equation}
\Sigma = a_1 \, a_2 \, \gamma \;,
\end{equation}
where $a_1$ and $a_2$ are the amplitudes of the LG10 modes (as given by Equation~\ref{eq:LG10_mag}). This quantity can be used to determine the effectiveness of pairs of actuators, as those with higher actuation ranges, or with lower degeneracy in actuation quadratures, subtend a larger area of phase space. It is our best metric for determining mode-matching capability. Table~\ref{Tab:area_matrix} lists the $\Sigma$-values for different pairings of proposed actuators  (see \S\ref{sec:actuator_designs}). Although the actuator area is computed in the space of the real and imaginary LG10 mode content, we can still visualize the region described by $\Sigma$ in the WS phase space to directly compare different pairs of actuators.

\begin{table}
\centering
\resizebox{\columnwidth}{!}{
\begin{tabular}{ | l | c | c | c | c| c |}
    \hline
&  OM2&  OM1&  FI&  SRM&  SR3
  \\ \hline\hline
$S$ (diopters)  & 1.4e-01 & 1.4e-01 & -5.0e-02 & 5.0e-02 & 4.7e-05\\ \hline
$a$-value  & 1.1e-01 & 9.7e-02 & -3.3e-01 & 5.6e-01 & 2.9e-01\\ \hline \hline
    {\bf LG10 area, $\Sigma$ }  & & &  & & \\ \hline
OM2  & 0  & 0.03 & {\bf0.51} & {\bf0.83} & 0.30 \\ \hline
OM1  & 0.03 & 0  & {\bf0.43} & {\bf0.68} & 0.21 \\ \hline
FI  & {\bf0.51} & {\bf0.43} & 0  & 0.33 & {\bf0.74} \\ \hline
SRM  & {\bf0.83} & {\bf0.68} & 0.33 & 0  & {\bf1.00} \\ \hline
SR3  & 0.30 & 0.21 & {\bf0.74} & {\bf1.00} & 0  \\ \hline
    \hline
 \end{tabular}}
\caption{The actuation strength of each proposed actuator. The normalized area of LG10 phase space covered by different combinations of actuators is also shown.}
\label{Tab:area_matrix}
\end{table}


\section{Actuation strategy \& tolerance study}
\label{sec:act_strategy}

In this section, we explore combinations of actuators of the output mode-matching in Advanced LIGO. We first determine the range and relative phase of each actuator in WS space and, from this, infer the optimal combination of actuators. We then determine the possible starting location in WS space based on the design tolerances of the interferometer, using the current LIGO Livingston (LLO) optical system as an example, and apply these actuators to evaluate their mode-matching capability. This analysis illustrates where design changes are required.

\subsection{Optimal actuator combination}
\label{sec:optimal_combination}

The two-dimensional nature of WS phase space implies that at least two {\it non-degenerate} actuators are required to achieve optimum mode-matching. That is, in order to achieve maximum overlap with the OMC, we need to match both the size and defocus of the field exiting the interferometer to the OMC mode. As was shown in \S\ref{sec:orthogonality}, optimum non-degeneracy occurs when the Gouy phase separation of the two actuators is $45^{\circ}$, in which case there is $90^{\circ}$ between actuation phases (i.e., the actuators are orthogonal). Table~\ref{Tab:gamma_matrix} lists the degeneracies between the actuators discussed in \S\ref{sec:actuator_designs}.

Pairs of actuators define an area in phase space that is accessible when actuation is provided (see \S\ref{sec:gouy_phase}). This area incorporates the actuation strength and the relative phase of different actuators to provide a metric for the optimum mode-matching. Table~\ref{Tab:area_matrix} lists the relative area in LG10 phase space spanned by different pairings of actuators. Our goal is to provide the maximum area with the minimum number of actuators. The existing SR3 actuator is reserved for matching the SRC mode to the common ARM mode. Thus for correcting the mode-mismatch between the ARM and OMC modes, at least two new actuators are required.

\begin{figure}[t]
  \centering
  \includegraphics[width=\columnwidth]{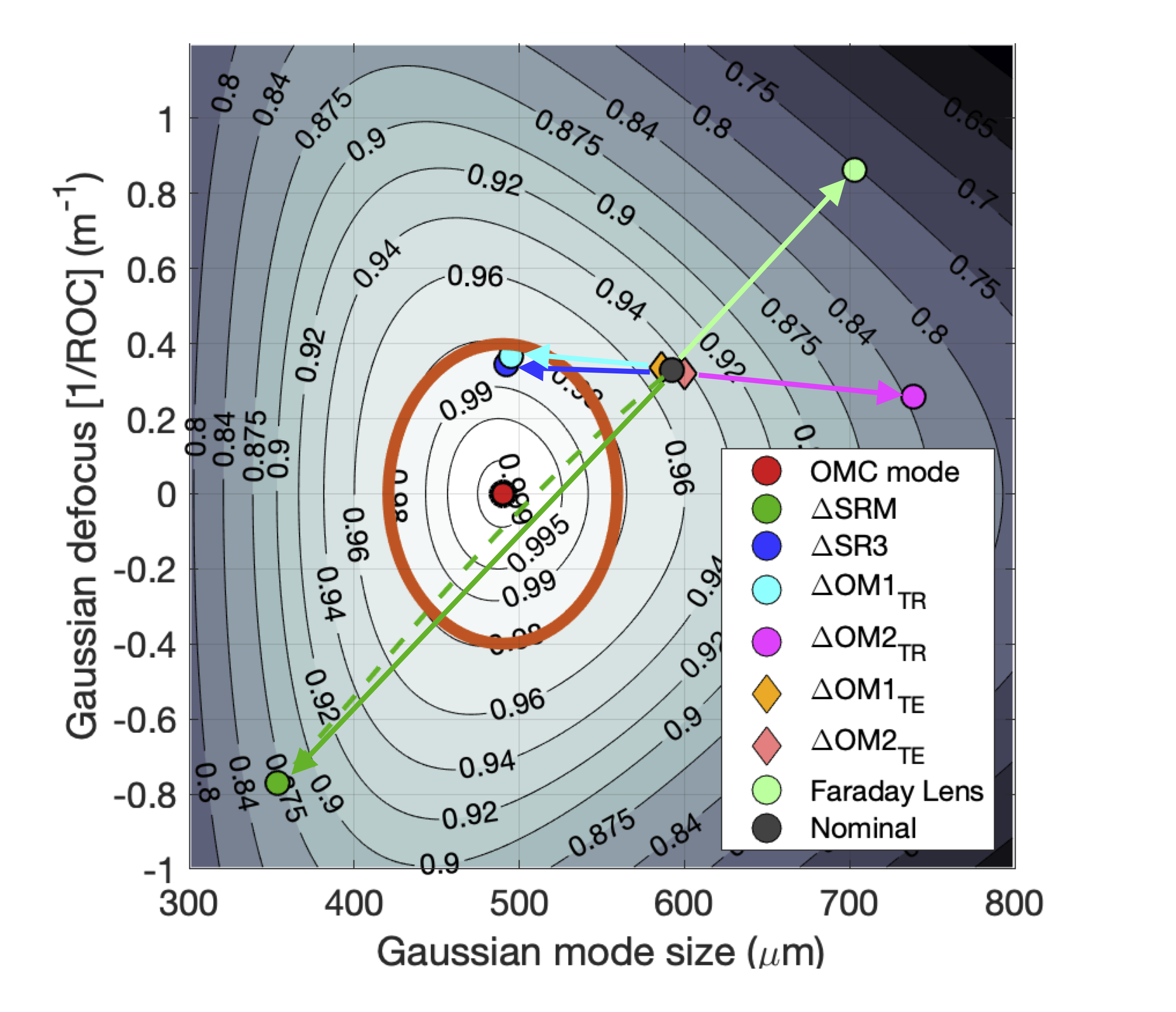}
  \caption{The effects of defocus actuators placed at different Gouy phases, propagated to the location of the OMC waist. In each case, the nominal ARM mode is displaced in WS space in response to a different actuation. The length of each vector represents the maximum displacement achievable by that actuator. This figure is explained in detail in \S\ref{sec:optimal_combination}.}
  \label{fig:MM_space_GP}
\end{figure}

Figure~\ref{fig:MM_space_GP} shows the effect of the actuators on the nominal ARM mode of the LLO interferometer (dark grey dot), which does not have 100\% overlap with the OMC mode (red dot). The target region of better than 98\% mode-overlap is enclosed by the orange line. In each case, the nominal ARM mode is displaced in WS space in response to a different actuation. The length of each vector represents the maximum displacement achievable by that actuator. For small actuation (as shown here), the orthogonality of different actuators can be directly inferred from the angle between their displacement vectors.

For larger actuation, nonlinear effects complicate this interpretation (see \S\ref{sec:gouy_phase}). We evaluate the significance of these effects for the actuator with the largest dynamic range, the SRM projector ($a=0.56$). The dashed green line in Figure~\ref{fig:MM_space_GP} shows the trajectory of the actuated mode through WS space as the actuation strength is increased from zero to maximum. It illustrates that, with increasing actuation strength, the true mode-actuation trajectories increasingly deviate from the linear trajectories indicated by the vector arrows. The linearized description is valid for actuation $\Delta S << 2 \lambda / \pi W^2$.

Overall, the effects of the actuators can be summarized as follows:
\begin{itemize}
\item The actuation range of the SRM substrate lens is the largest.
\item The SRM and FI lenses are approximately anti-symmetric with respect to each other.
\item The OM1 and OM2 actuators are roughly orthogonal to the SRM and FI lenses.
\item The OM1 and OM2 actuators are approximately anti-symmetric with respect to each other.
\item Thermo-elastic actuation of OM1 and OM2 is ineffectual, but thermo-refractive actuation is comparable in strength to the SRM lens.
\end{itemize}
Therefore, a combination of the thermo-refractive versions of the OM1 and OM2 actuators, in conjunction with the SRM and FI substrate lenses, will be able to access a {\it significantly} larger region of phase space than would a single actuator.

\subsection{Mode-matching capability accounting for real-world design tolerances}
\label{sec:MMopt}


\begin{table*}
\centering
{
\begin{tabular}{|l||r|r|}
  \hline
   {\bf Parameter Name} & {\bf LLO Value} & {\bf Uncertainty} \\ \hline \hline
   {\bf ARMs} & & \\ \hline
   ITM ROC & 1939 m & $\pm$6.0 m \\ \hline
   ETM ROC & 2240 m & $\pm$6.4 m \\ \hline
   ARM length & 3995 m & $\pm$3 mm \\ \hline \hline
   {\bf SRC} & & \\ \hline
   SRM ROC & -5.678 m & $\pm$0 mm \\ \hline
   SR2 ROC & -6.425 m & $\pm$6 mm \\ \hline
   SR3 ROC & 36.013 m & $\pm$36 mm \\ \hline
   ITM-SR3 length (common) & 24.368 m & $\pm$3 mm \\ \hline
   SR3-SR2 length & 15.461 m & $\pm$3 mm \\ \hline
   SR2-SRM length & 15.739 m & $\pm$3 mm \\ \hline
   ITM static substrate lens (common) & -2.35 $\mu$D & $\pm$0 $\mu$D \\ \hline
   CP lens (common) & 28 $\mu$D & $\pm$0 $\mu$D \\ \hline \hline
   {\bf Output optics} & & \\ \hline
   OM1 ROC & 4.60 m & $\pm$35 mm \\ \hline
   OM2 ROC & 1.70 m & $\pm$90 mm \\ \hline
   OM3 ROC & flat & $\pm$0 mm \\ \hline
   SRM (AR surface)-OM1 length & 3.410 m & $\pm$3 mm \\ \hline
   OM1-OM2 length & 1.390 m & $\pm$3 mm \\ \hline
   OM2-OM3 length & 0.640 m & $\pm$3 mm \\ \hline
   OM3-OMC (waist) length & 0.450 m  & $\pm$3 mm \\ \hline
   SRM static substrate lens & 79 mD & $\pm$0 mD \\ \hline
\end{tabular}}
\caption{Output-side optical parameters of the Advanced LIGO Livingston (LLO) detector and their design tolerances.}
\label{Tab:IFO_parameters}
\end{table*}

With a set of actuators identified, we next ask: Given the design tolerances on all distances and radii of curvature, what is the probable starting region in WS space, and what actuation ranges are required to maximize the overlap of the ARM, SRC, and OMC modes?

The analysis in this section is performed using the Finesse interferometer modeling software \cite{FinesseRef}. With Finesse, we can model the effect of actuators on resonant cavity modes, as is necessary for considering actuation of optics inside the SRC. We present an analysis of the LLO inteferometer as a case study of this technique. The parameters and tolerances for the LLO optical system are given in Table~\ref{Tab:IFO_parameters}.

\begin{figure}[ht]
  \centering
  \includegraphics[width=\columnwidth]{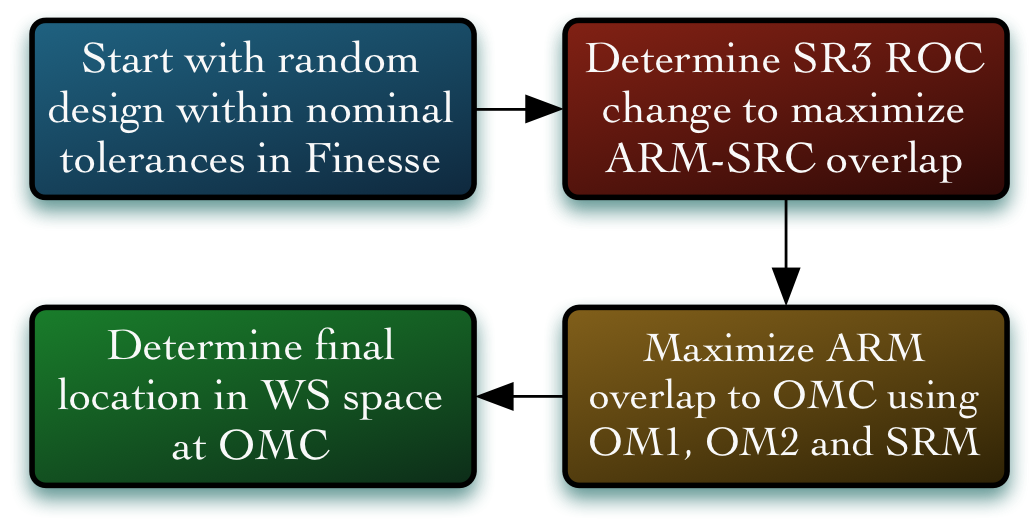}
  \caption{Overview of the Finesse procedure for maximizing the output mode-matching, given a set of design tolerances.}
  \label{fig:finesse_procedure}
\end{figure}

\begin{figure*}[t]
  \begin{minipage}[l]{\columnwidth}
    \centering
    \includegraphics[width=0.95\columnwidth]{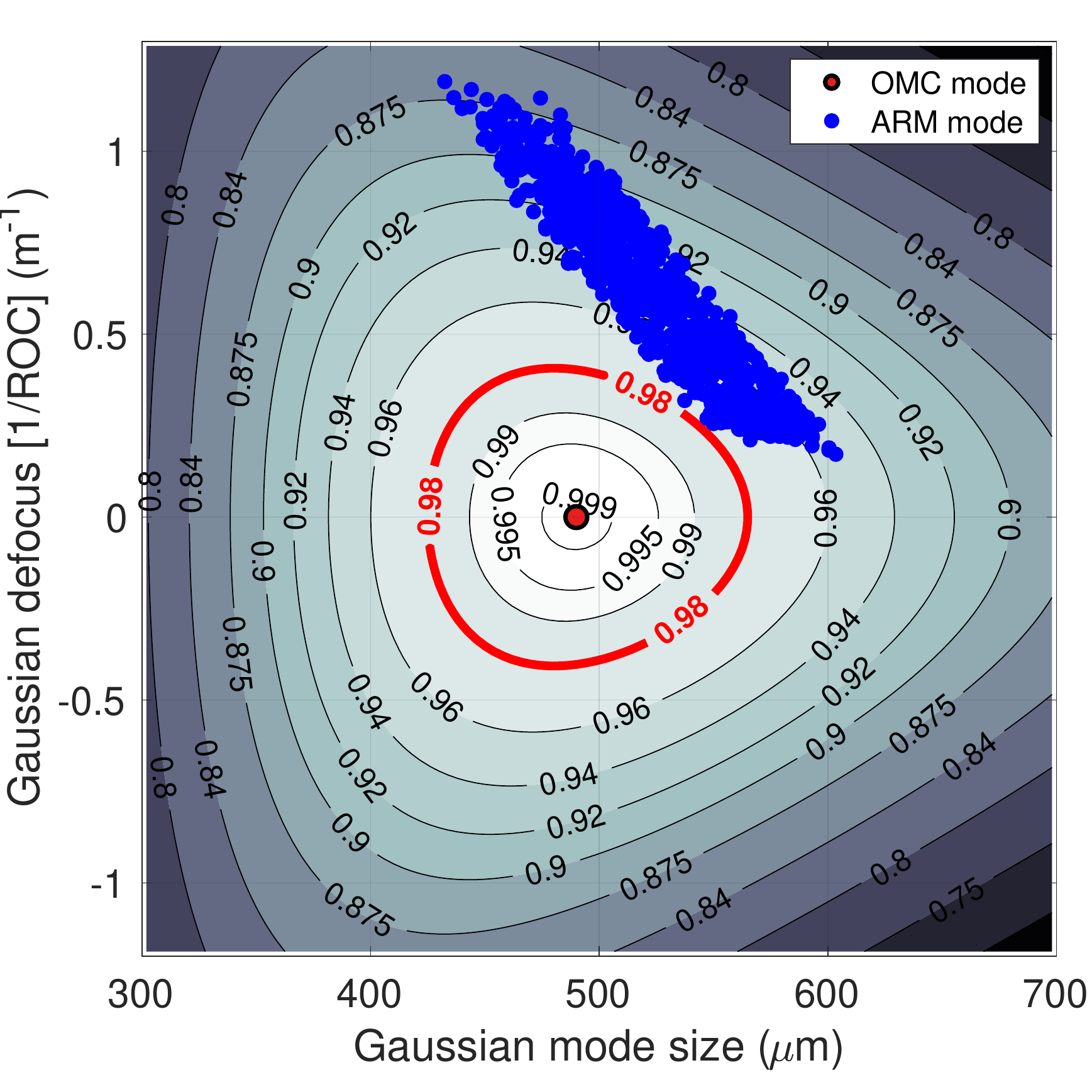}
    (a) Initial region, before any actuation is applied.
  \end{minipage}
  \hfill{}
  \begin{minipage}[r]{\columnwidth}
    \centering
    \includegraphics[width=0.95\columnwidth]{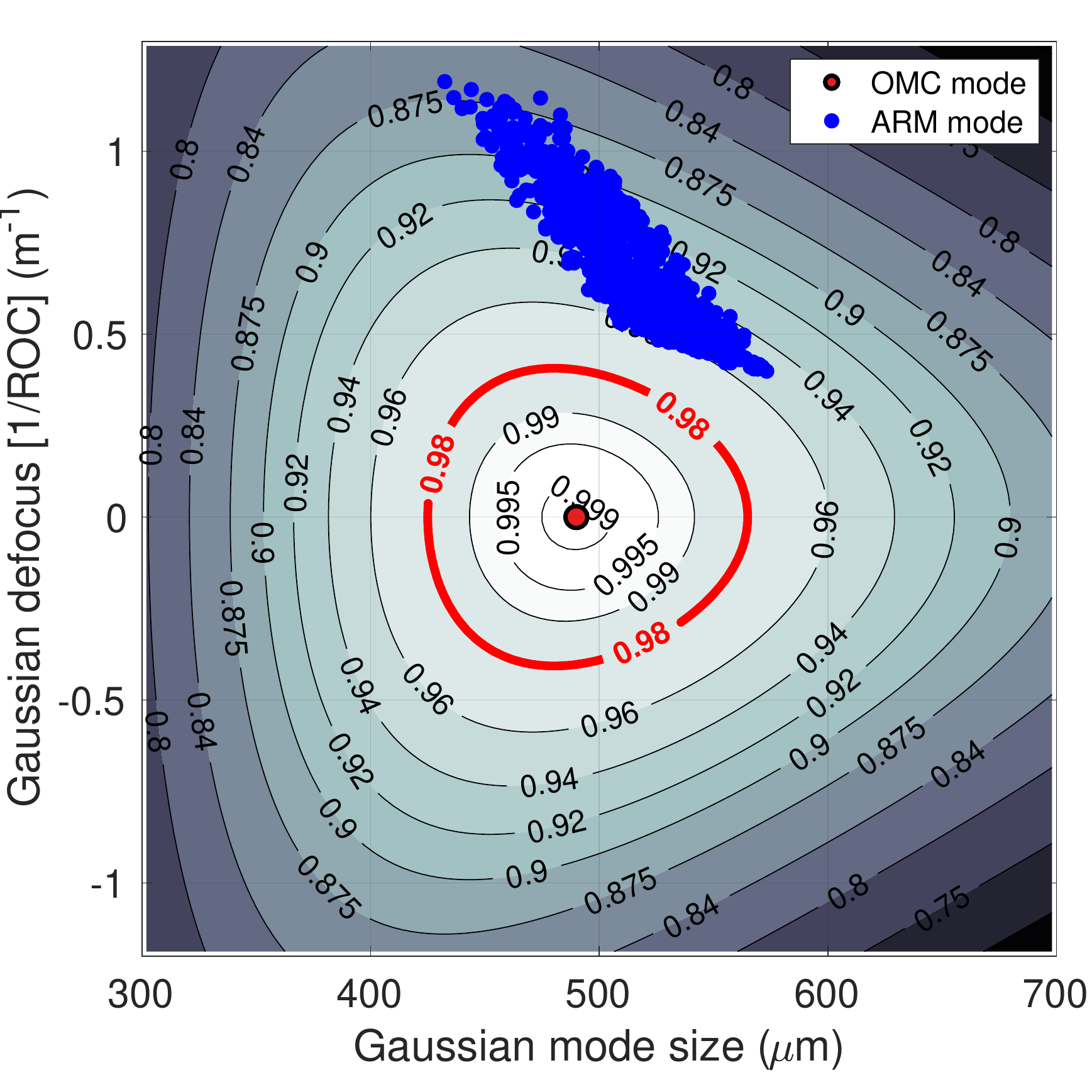}
    (b) After actuation of SR3 (optimal SRC-ARM matching).
  \end{minipage}
  \par\bigskip
  \begin{minipage}[l]{\columnwidth}
    \centering
    \includegraphics[width=0.95\columnwidth]{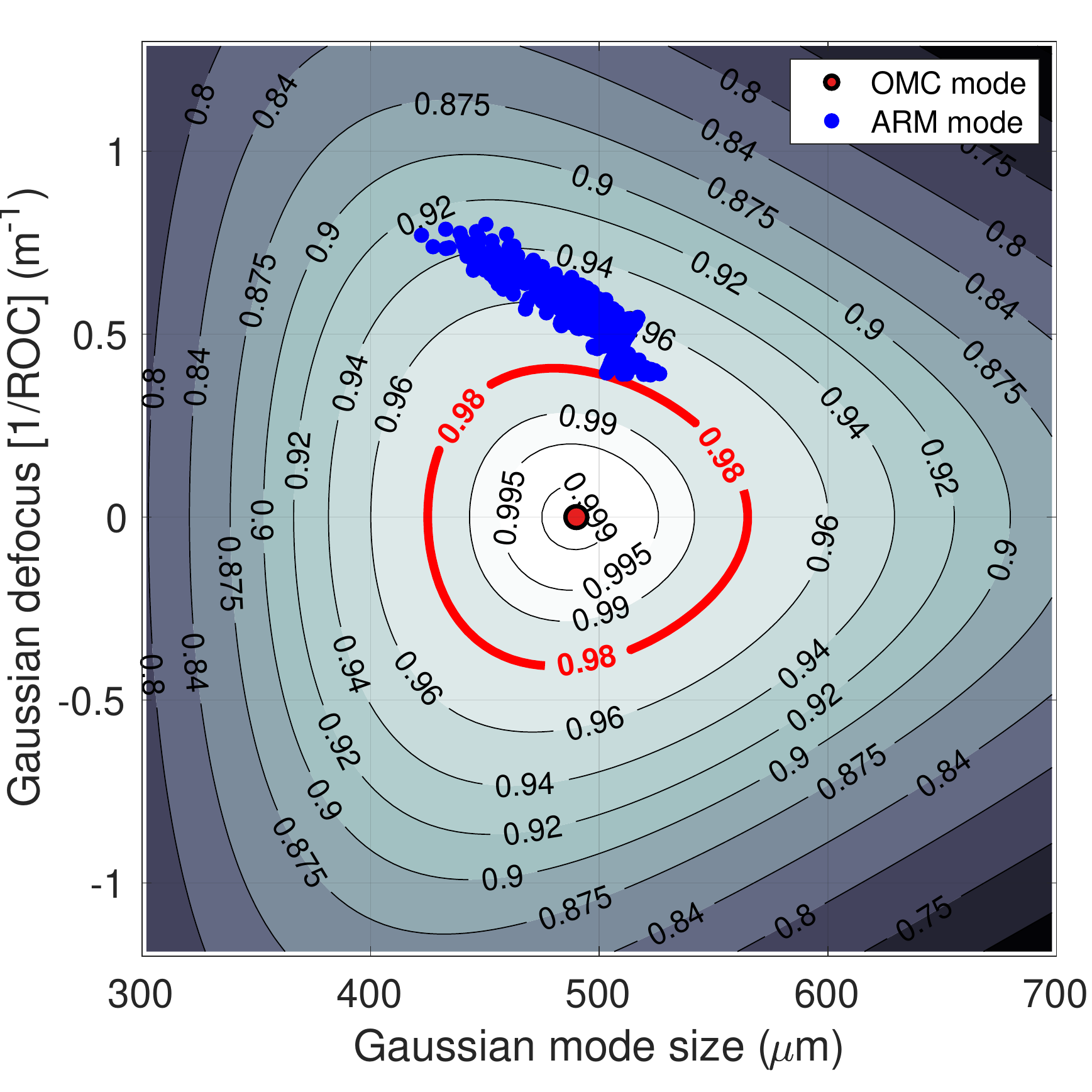}
    (c) After actuation of SR3, OM1, and OM2.
  \end{minipage}
  \hfill{}
  \begin{minipage}[r]{\columnwidth}
    \centering
    \includegraphics[width=0.95\columnwidth]{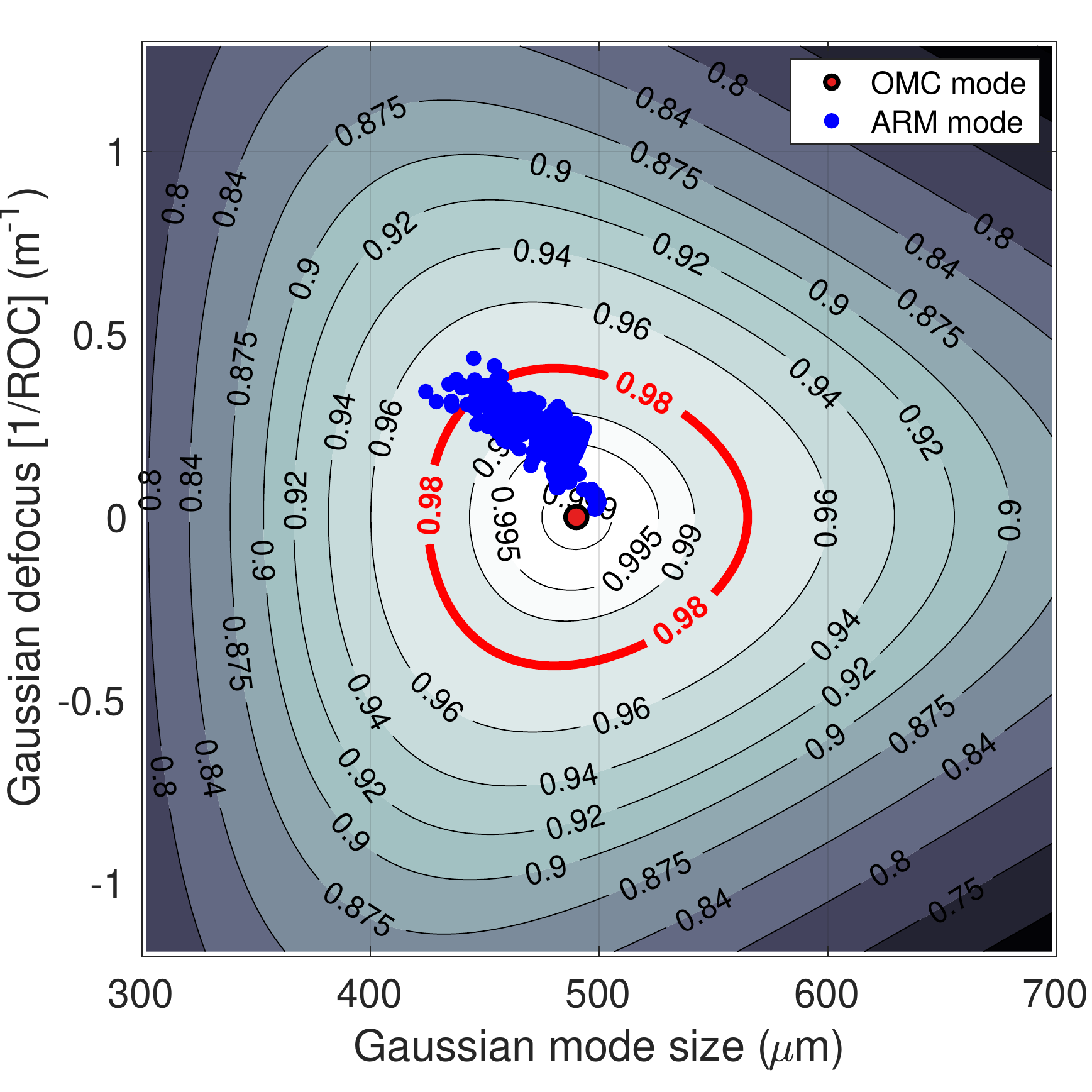}
    (d) After actuation of SR3, OM1, OM2, SRM, and FI.
  \end{minipage}
  \caption{Region of phase space expected to contain the ARM mode, given the uncertainties in the radii of curvature and distances between optics. Each panel shows the ARM modes of 1,000 randomized configurations within the Advanced LIGO design tolerances, propagated to the location of the OMC waist. (a): The initial possible region before any actuation is applied. (b): The expected region after optimal actuation of the SR3 mirror within its allowed range. (c): The expected region after optimal actuation of the SR3, OM1, and OM2 mirrors within their allowed ranges. (d): The final expected region after optimal actuation of the SR3, OM1, and OM2 mirrors and the SRM and FI substrate lenses within their allowed ranges. The locations of the final ARM modes confirm that the proposed actuation strategy can achieve less than $2\%$ mean output mode-matching loss.}
  \label{fig:MM_finesse}
\end{figure*}

Our procedure for optimizing the mode-matching to the OMC is illustrated in Figure~\ref{fig:finesse_procedure}. In more detail, the analysis entails the following steps:
\begin{enumerate}
  \item Start with all nominal distances and radii of curvature in the optical layout. For each value, add an error drawn from a uniform distribution within the design tolerances (see Table~\ref{Tab:IFO_parameters}) to create a randomized parameter set.
  \item For this parameter set, run Finesse to solve for the initial ARM and SRC modes and propagate them to the location of the OMC waist.
  \item Repeat this procedure for 1,000 randomized parameter sets.
\end{enumerate}
This allows us to determine the initial distributions of ARM and SRC modes in WS space at the OMC, as shown in Panel~(a) of Figure~\ref{fig:MM_finesse}. The assumption of uniform parameter distributions is conservative: a normal distribution would de-weight the probability of values near the edge of the tolerance range, but here we assume only that each value is somewhere ``within specification.'' 1,000 simulations are found to adequately sample the parameter space, as they fully resolve the sharp bounds enforced by these uniform priors (visible as the sharp edges of the distribution in Panel~(a)).

At this point, we can take full advantage of the Finesse simulation by actuating the SR3 ROC within its allowed range to improve the ARM-SRC mode-matching. That is, for each of the 1,000 randomized parameter sets, the resonant TEM00 eigenmodes of the ARM cavities and SRC are continually solved in Finesse as the SR3 ROC is adjusted. For each SR3 ROC value, the beam size and defocus of the ARM and SRC modes are calculated at the same longitudinal plane (in our case, at the OMC waist) and their overlap is evaluated using Equation~\ref{eqn:OL}. Finally, the SR3 ROC is set to the value which maximizes this overlap. Maximizing the mode-overlap between the ARM cavities and SRC eliminates losses in the signal recycling path and drives the interferometer frequency response (coupled cavity pole) as close as possible to its theoretical value. Panel~(b) of Figure~\ref{fig:MM_finesse} shows the change in the distributions of ARM and SRC modes after the optimal SR3 actuation is applied to each parameter set.

The Finesse procedure then continues as follows:
\begin{enumerate}
  \setcounter{enumi}{3}
    \item For each randomized parameter set, optimize the actuation of SR3 within its allowed range to maximize the overlap of the ARM and SRC modes, as described above.
    \item For each parameter set, analogously optimize the actuation of OM1, OM2, and the SRM and FI substrate lenses to maximize the overlap of the ARM and OMC modes.
\end{enumerate}
The effect of the final step is shown in the bottom two panels of Figure~\ref{fig:MM_finesse}. Panel (c) shows the achievable mode-matching when the OM1 and OM2 actuators are implemented, and Panel (d) shows the improvement when the SRM and FI substrate lenses are additionally included. This analysis confirms that, given the design tolerances of the Advanced LIGO interferometers, the proposed actuation strategy can achieve less than $2\%$ mean output mode-matching loss.


\section{Conclusions}
\label{sec:conclusion}

In conclusion, achieving $-10$~dB of squeezing in Advanced LIGO will require reducing the output mode-matching losses to less than $2\%$. We have shown that this will require additional defocus actuators and/or a redesign of the SRC-OMC output chain. We have introduced a phase space, WS-space, which is effective at visualizing and understanding the relationship between different interferometer modes, the subsquent mismatch between them, and the effect of different mode-matching actuators on those modes.

In a case study of the current LLO optical system, we used a statistical approach with randomized starting configurations  (determined by variations of the distances and radii of curvature of the interferometer optics from their nominal values) to visualize the distribution of possible modes within WS-space. The total output mode-matching loss  varies from 15\% to a few percent for the different randomized configurations. The existing SR3 actuator is required to improve losses between SRC and the ARM cavities. This improves each configuration in overlap between the SRC and ARM modes, but not necessarily in total losses. Complete correction is achieved with use of four optics external to the SRC, which correct for losses between the whole interferometer and the OMC.

Over all random configurations, the total correction requires a maximum actuation of $+50$~mD on the SRM substrate and $-50$~mD on the introduced new transmissive optic, FI. The OM1 and OM2 mirrors each require a maximum actuation of $+140$~mD. On SR3 the current actuation range of $+47\;\mu$D is assumed. Due to their anti-symmetry, the two substrate lenses (or the two OMs) can be used differentially to achieve a combined range of $\pm100$~mD ($\pm 280$~mD). These requirements are within the demonstrated ranges of current similar actuators, and thus are feasible with existing technology. The significant improvement for all cases is very clearly shown in the WS-space, demonstrating the efficacy of this new visualization.


\section*{Acknowledgments}
We are grateful to Evan Hall for providing the O3 noise curves, John Miller for providing the squeezing loss code, and Hiro Yamamoto for helpful comments in the early stages of this project. We also thank Daniel Brown for helpful comments during LSC review. LIGO was constructed by the California Institute of Technology and Massachusetts Institute of Technology with funding from the National Science Foundation and operates under cooperative agreement PHY-0757058. This paper has LIGO Document Number LIGO-P1900375.

\appendix

\section{Actuator designs}
\label{sec:actuators_physical}

In this appendix, we consider devices capable of actuating on the interferometer modes. Our general requirements for an ideal wavefront actuator are: (a) large dynamic range, (b) low displacement noise, (c) high-quality  wavefront correction (i.e., low spatial distortion upon correction), and (d) low backscatter. The following section discusses real actuators that have been demonstrated in similar or identical circumstances and configurations to the proposed implementation. We consider only actuators that can be applied to the existing infrastructure \cite{TipTilt:10} without substantial redesign (i.e., do not require new suspended large optics or significant topological changes).

In order to avoid damaging the optics during actuation, we set limits on the maximum stress and temperature allowed in the optic. We set the maximum stress to \SI{5}{\mega\pascal}, approximately 10\% of the bending strength and tensile strength of fused silica \cite{heraeus_fused_silica}. Dielectric coatings are typically annealed at \SI{400}{\celsius} - \SI{500}{\celsius}. To avoid exceeding about 20\% of this temperature, we specify a maximum permissible $\Delta T$ of \SI{100}{\kelvin}, equal to a maximum temperature of roughly \SI{120}{\celsius}. This assumes a safety factor of approximately $4\times$ for the temperature. Less conservative operation will, of course, extend the range of these actuators.

\begin{figure}[t]
  \centering
  \includegraphics[width=\columnwidth]{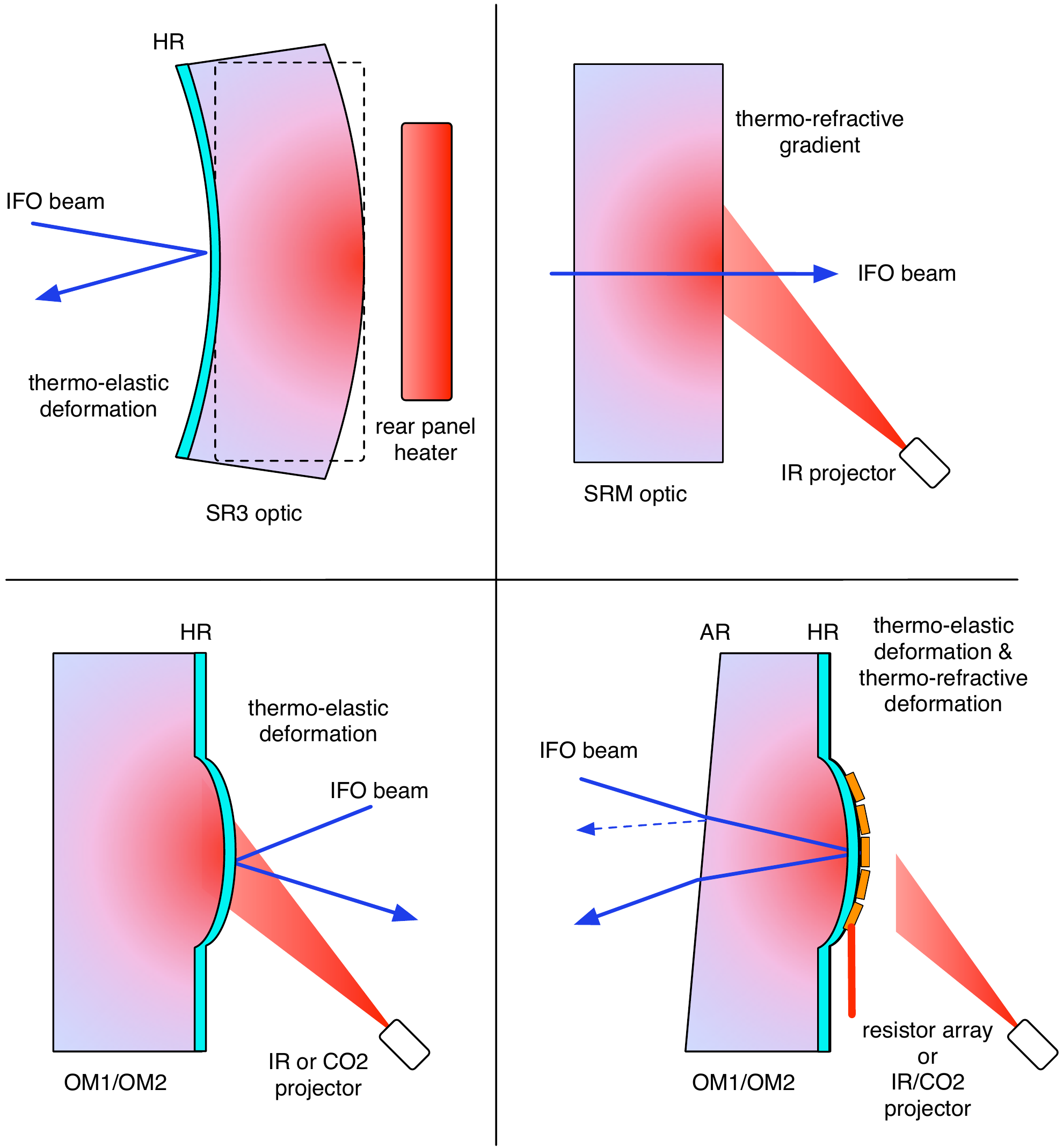}
  \caption{Examples of possible wavefront actuators. Top left: tunable thermo-elastic surface curvature change of SR3 optic. Top right: tunable thermo-refractive lens in the substrate of the SRM. Bottom left: tunable thermo-elastic lens in the surface of OM1 or OM2. Bottom right: tunable thermo-refractive lens in the substrate of OM1 or OM2.}
  \label{fig:actuators}
\end{figure}

\subsubsection{SR3 heater}

The SR3 heater is an existing actuator which heats the back surface of SR3, as illustrated in the top left panel of Figure~\ref{fig:actuators}. It has been found to produce a change in surface curvature of approximately \SI[per-mode=symbol]{-3.05}{\milli\meter\per\watt} \cite{SR3_heater_aLOG} in the case where SR3 is a concave mirror with a ROC of \SI{36}{\meter}. The existing electrical implementation of this actuator is limited to approximately 10~W, allowing the mirror curvature to be reduced by up to \SI{30.5}{\milli\meter}. The maximum defocus change for the reflected beam is \SI{47}{\micro\diopter}. The SR3 heater actuates on the SRC mode, affecting the matching of the ARM mode to the SRC and OMC modes.

\subsubsection{SRM substrate lens}
\label{sec:SRM_substrate}

The SRM substrate actuator is a proposed design that would introduce a thermal lens within the substrate of the SRM (outside the SRC) via a $\rm CO_2$ laser beam incident on the {\it back surface} of the optic. This is illustrated in the top right panel of Figure~\ref{fig:actuators}. We consider a \SI{750}{mW} $\rm CO_2$ laser with a beam diameter at the SRM of \SI{8}{\milli\meter}, approximately twice the interferometer beam diameter. Conceptually, this is the same as the $\rm CO_2$ central heating of the compensation plates used in Advanced LIGO \cite{Brooks:16} and the adaptive optic element described in detail by Arain \cite{Arain:07}.

The lens strength can be approximated with the formula for the coating-induced absorption sagitta of a wavefront from Winkler et al. \cite{PhysRevA.44.7022}. In this case, the sagitta is the optical path length difference at one heating beam radius, $w$. For thermo-elastic deformation on transmission through an optic, the sagitta is

\begin{equation}
ds =  \frac{n\,\alpha P}{4\, \pi \kappa},
\end{equation}

\noindent where $n$ is the refractive index of the optic (1.45 for fused silica), $\alpha$ is the coefficient of thermal expansion (\SI{0.55e-6}{\per\kelvin}), $P$ is the absorbed power and $\kappa$ is the thermal conductivity (\SI{1.38}{\watt\per\meter\per\kelvin}). The thermo-refractive sagitta is

\begin{equation}
ds =  \frac{\beta P}{4\, \pi \kappa},
\end{equation}

\noindent where $\beta$ is the thermo-optic coefficient (\SI{8.6e-6}{\per\kelvin}).

The defocus ($S$) of a wavefront profile ($U$) is represented by the coefficient of the quadratic term of a wavefront,

\begin{equation}
    U = \frac{1}{2}\,S\,r^2.
\end{equation}

\noindent As the sagitta equals the wavefront at $r=w$, the defocus can be expressed as

\begin{equation}
S = 2\, \frac{ds}{w^2},
\end{equation}

\noindent and thus the total lens strength, $S_{SRM}$, is given by

\begin{equation}
S_{SRM} = \frac{\left(\beta + n\, \alpha \right) P}{2\, \pi \kappa w^2}.
\end{equation}

\noindent For the $\rm CO_2$ laser source described above, this yields approximately \SI{50}{\milli\diopter}. Note that the thermo-elastic effect is approximately 6\% of the size of the thermo-refractive effect. The induced lens will affect the mode-matching of all modes relative to the OMC mode.

We note that the assumption of \SI{750}{mW} of delivered $\rm CO_2$ laser power is conservative and the power could be increased, if required. For a $\rm CO_2$ laser source of double the power (\SI{1.5}{\watt}), finite-element modeling of the mirror shows a maximum temperature of approximately \SI{110}{\kelvin} above room-temperature (or \SI{130}{\celsius}, assuming a room temperature of \SI{20}{\celsius}) and a peak von Mises stress of \SI{2}{\mega\pascal}. These are still safely within the limits for fused silica.

\subsubsection{FI substrate lens}
\label{sec:FI_substrate}

An alternative to the SRM actuator is a new transmissive optic between the SRM and the OMC, mounted to the OFI assembly. We will refer to this new optic as FI. It offers several advantages. First, thermal actuation can be provided more simply using an annular heating ring around the outside of the optic, as described by Arain \cite{Arain2010UFLens}. This option eliminates the need for a new $\rm CO_2$ laser source and its accompanying alignment considerations. Second, while the SRM actuation is unidirectional, it is possible to invert the sign of the FI actuation by incorporating a static (unheated) ROC offset in the lens, which is {\it reduced} as heat is applied. Throughout, we will assume that the FI actuator is used in this way to provide opposite-signed actuation, compared to the SRM actuator.

\subsubsection{OM1 and OM2 heaters}

The OM1 and OM2 heaters are proposed actuators to introduce two additional, independently-tunable thermal lenses between the SRC and the OMC. We consider two different designs, illustrated in the bottom two panels of Figure~\ref{fig:actuators} and each described below.

First, by heating the front surface of OM1 and OM2 with an infrared heater beam or a $\rm CO_2$ laser beam, as illustrated in the bottom left panel of Figure~\ref{fig:actuators}, we can create a localized surface deformation that approximates a change in the local ROC. This is conceptually the same as the CHRoCC system used in the Virgo gravitational wave detector \cite{CHRoCC:2013etal} and the adaptive optic element described by Arain \cite{Arain:07}. The approximate defocus added to the interferometer laser beam upon reflection (due solely to the thermo-elastic effect shown in the lower left panel of Figure \ref{fig:actuators}) is

\begin{equation}
S_{OM_{TE}} = -\frac{\alpha\, P}{\pi \kappa w^2},
\end{equation}

\noindent where the parameters are defined as in \S\ref{sec:SRM_substrate}. Note that a factor of $2$ has been added here to account for the double-pass effect that occurs with reflection relative to transmission. With a \SI{570}{\milli\watt} laser and a \SI{3}{\milli\meter} diameter spot size, we would be limited to approximately \SI{8.7}{\milli\diopter} of actuation range. Under these conditions, the peak temperature in the optic would be approximately \SI{100}{\kelvin} above ambient and the peak stress would be approximately \SI{2}{\mega\pascal}. In this case, we have limited the delivered laser power so that the peak temperature does not exceed \SI{120}{\celsius}.

Alternatively, we can use the OM1 and OM2 mirrors in the configuration illustrated in the lower right panel of Figure~\ref{fig:actuators} in which the highly-reflective (HR) coating is applied to the {\it back surface} of the optics. In this configuration, the interferometer beam double-passes the substrate when reflecting off the HR surface, so additionally undergoes thermo-refractive lensing. The approximate defocus added to the interferometer beam from the combined thermo-optic and thermo-elastic effects is

\begin{equation}
S_{OM_{TR}} = \frac{\left(\beta + n\, \alpha \right) P}{ \pi \kappa w^2}.
\end{equation}

\noindent For a \SI{570}{\milli\watt} laser, the effective lens is approximately \SI{145}{\milli\diopter}. This design is simply a variant of the Advanced LIGO $\rm CO_2$ central heating case \cite{Brooks:16}. A similar configuration with resistive heater elements bonded to the back surface of the optic (instead of $\rm CO_2$ laser heating), also illustrated in the bottom right panel of Figure~\ref{fig:actuators}), has been demonstrated by Kasprzack \cite{Kasprzack:13}. Under either design, the OM1 and OM2 lenses will affect the mode-matching of all modes relative to the OMC mode.


\bibliography{GWreferences2}
\bibliographystyle{ieeetr}

\end{document}